\documentclass[prl,twocolumn,preprintnumbers,superscriptaddress]{revtex4}

\usepackage{graphicx}
\usepackage{dcolumn}
\usepackage{amsmath}
\usepackage{color}

\begin{document}

\title{Spin-density, charge- and bond-disproportionation wave instability in hole-doped infinite-layer $R$NiO$_2$}

\author{K. G. Slobodchikov}
\affiliation{Ural Federal University, 620002 Yekaterinburg, Russia}

\author{I. V. Leonov}
\affiliation{M. N. Miheev Institute of Metal Physics, Russian Academy of Sciences, 620108 Yekaterinburg, Russia}
\affiliation{Ural Federal University, 620002 Yekaterinburg, Russia}

\begin{abstract}
Using \emph{ab initio} band structure methods and DFT+dynamical mean-field theory approach we explore the possible formation of spin and charge stripes in the Ni-O plane of hole-doped infinite-layer nickelates, $R$NiO$_2$.  Our results reveal a remarkable instability of the $C$-type $(110)$ spin state with undistorted lattice towards the formation of the spin-density, charge- and bond-disproportionation stripe phases accompanied by in-plane``breathing''-like distortions of the crystal structure. Our work gives a comprehensive picture of competing charge and spin stripe states, with possible frustration of different stripe patterns upon doping. It suggests that the spin and charge stripe state likely arises from strong magnetic correlations (with concomitant lattice distortions), which play a key role for understanding the anomalous properties of hole-doped layered nickelates.
\end{abstract}

\maketitle


The recent discovery of unconventional superconductivity in hole-doped  infinite-layer nickelates ($R$NiO$_2$ with $R$ = rare-earth element) which depending upon composition, doping, and pressure show superconductivity below $T_c \sim31$~K has garnered significant research interest around the world \cite{Li_2019,Hepting_2020,Zeng_2020,Osada_2021,Goodge_2021,
Lu_2021,Wang_2021,Pan_2022,Zeng_2022,Kitatani_2020,Chen_2022a,
Nomura_2022,Gu_2022,Botana_2022}. $R$NiO$_2$ crystallizes in an ``infinite-layer'' planar crystal structure similar to that of the parent hole-doped superconductor CaCuO$_2$ with a critical temperature up to $\sim$110 K. In $R$NiO$_2$ Ni ions adopt a nominal Ni$^+$ $3d^9$ configuration (with the planar Ni $x^2 - y^2$ orbital states dominated near the Fermi level) being isoelectronic to Cu$^{2+}$ in CaCuO$_2$ \cite{Azuma_1992,Peng_2017,Savrasov_1996}. Despite this apparent similarity the low-energy physics of hole-doped $R$NiO$_2$ exhibits notable differences, e.g., the Ni $x^2 - y^2$ states are found to experience strong hybridization with the rare-earth $5d$ orbitals, yielding a noncuprate-like (multi-orbital) Fermi surface \cite{Anisimov_1999,Lee_2004,Botana_2020,Choi_2020}. In addition, experimental and theoretical estimates suggest a relatively large charge-transfer energy in $R$NiO$_2$ \cite{Hepting_2020,Goodge_2021}. This implies that the electronic structure of $R$NiO$_2$ is close to a Mott-Hubbard regime, distinct from a charge-transfer state in superconducting cuprates. The former also highlights the crucial importance of strong electronic correlations \cite{Mott_1990,Imada_1998,Tokura_2000} to explain the properties of $R$NiO$_2$, consistent with the results of previous many-body DFT+dynamical mean-field theory (DFT+DMFT) \cite{Georges_1996,Kotliar_2006} and GW+DMFT \cite{Sun_2002,Biermann_2003} electronic structure calculations \cite{Werner_2020,Lechermann_2020a,Karp_2020a,Karp_2020b,
Lechermann_2020b,Wang_2020,Nomura_2020,Si_2020,Leonov_2020,
Ryee_2020,Lechermann_2021,Wan_2021,Leonov_2021,Kutepov_2021,Lechermann_2022,
Malyi_2022}. The DFT/GW+DMFT calculations show a remarkable orbital-dependent localization of the Ni $3d$ states, complicated by large hybridization with the rare-earth $5d$ states (while the rare-earth $4f$ states locate far away from the Fermi level due to the large Hubbard $U$ coupling). Moreover, it was shown that $R$NiO$_2$ undergoes a Lifshitz transition of the Fermi surface accompanied by a drastic change of magnetic correlations upon doping \cite{Leonov_2020,Leonov_2021}, implying a complex low-energy physics of infinite-layer nickelates.

While the magnetism of hole-doped $R$NiO$_2$ still remains debated \cite{Lin_2021,Zhou_2022,Lin_2022}, recent resonant inelastic x-ray scattering (RIXS) experiments on hole-doped $R$NiO$_2$ grown on and capped with SrTiO$_3$ reveal the existence of a sizable antiferromagnetic (AFM) correlations with dispersive magnetic excitations with a bandwidth $\sim$200~meV \cite{Lu_2021} consistent with a Mott system being in the strong coupling regime \cite{Mott_1990,Imada_1998,Tokura_2000}. Most interestingly, a translational symmetry broken state with a propagating wave vector $(0.33,0)$ r.l.u. (along the Ni-O bond) has been recently reported independently by different experimental groups, based on the RIXS near Ni $L_3$ absorption edge experiments for the uncapped hole-doped $R$NiO$_2$ grown on SrTiO$_3$ \cite{Rossi_2021,Tam_2021,Krieger_2021}. This suggests the formation of a superstructure of the lattice which has been naturally ascribed to the emergence of a charge-density wave instability (charge stripes), which seems to be a key ingredient for superconducting cuprates \cite{Tranquada_1995,Wells_1997,Salkola_1996,Keimer_2015,Huang_2022,Xiao_2022} as well as a characteristic feature of the hole-doped nickelates \cite{Lee_1997,Yoshizawa_2000,Botana_2016,Zhanga_2016,Bernal_2019,
Zhang_2019,Zhang_2020,Hao_2021}. In fact, the charge-density wave formation was discussed in the case of (La,Sr)$_2$NiO$_4$ (with Sr $x=1/3$, Ni$^{2.33+}$) \cite{Lee_1997,Yoshizawa_2000} as well as for the square-planar systems La$_4$Ni$_3$O$_8$ (Ni$^{1.33+}$) and La$_3$Ni$_2$O$_6$ (Ni$^{1.5+}$) \cite{Botana_2016,Zhanga_2016,Bernal_2019,Zhang_2019,Zhang_2020,Hao_2021}.
This raises the question about the mechanism of superconductivity and the role of spin and charge stripe fluctuations in the infinite-layer $R$NiO$_2$.


In this work, using the DFT+U \cite{Anisimov_1991,Liechtenstein_1995,Dudarev_1998,Giannozzi_2009,Giannozzi_2017} and DFT+DMFT \cite{Georges_1996,Kotliar_2006} electronic structure methods we explore the possible formation of spin and charge stripes in the Ni-O plane of hole-doped infinite-layer nickelates, $R$NiO$_2$.  Our results reveal an emergent instability of the $C$-type AFM spin state (with a magnetic vector $q_m=(110)$ at the Brillouin zone $M$ point) of hole-doped $R$NiO$_2$ with undistorted lattice towards the formation of the spin and charge stripe phases accompanied by in-plane``breathing''-like distortions of the crystal structure, with a possible frustration of different spin and charge stripe patterns at large doping. Our results provide a microscopic evidence of competing charge and spin stripe states, which seem to play a key role for understanding the anomalous properties of hole-doped nickelates.

\begin{figure}[tbp!]
\centerline{\includegraphics[width=0.5\textwidth,clip=true]{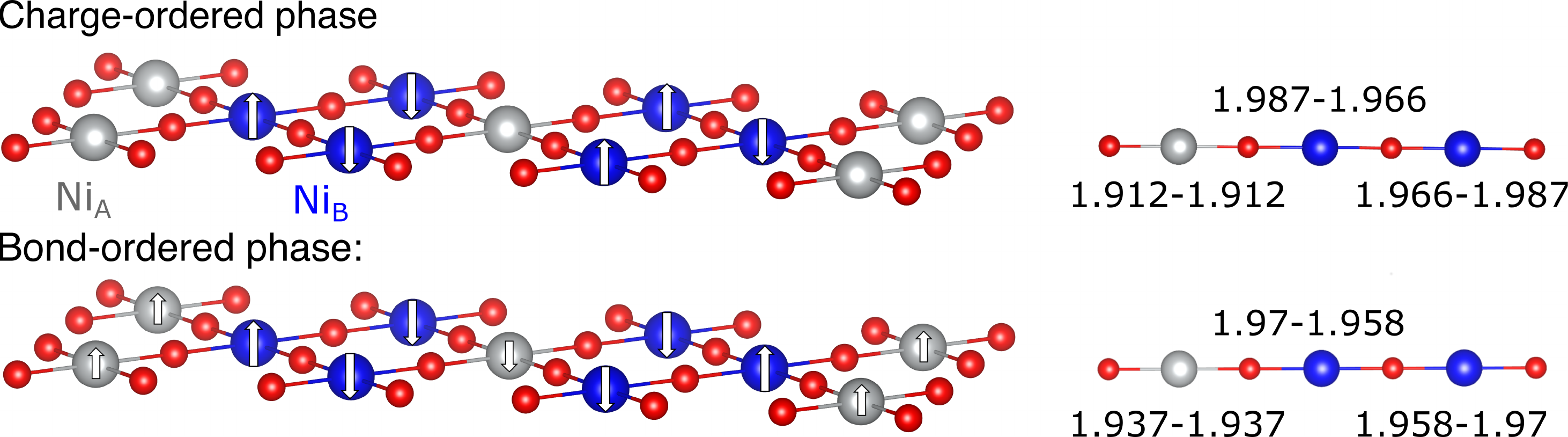}}
\caption{Left panel: proposed spin and charge ordering pattern inside the Ni-O plane in the charge-ordered (top) and bond-ordered phases (bottom) of hole-doped $R$NiO$_2$, with charge deficient ``Ni$^{2+}$'' ions (Ni$_\mathrm{A}$) shown in grey and nominal Ni$^{+}$ ions (Ni$_\mathrm{B}$) in blue. Arrows correspond to up/down spins for the Ni ions, with $\sim$0.68$\mu_\mathrm{B}$ spin moment for the Ni$_\mathrm{B}$ ions in the CO (Ni$_\mathrm{A}$ spin moment is zero) and $\sim$0.56$\mu_\mathrm{B}$ and 0.66$\mu_\mathrm{B}$ for the Ni$_\mathrm{A}$ and Ni$_\mathrm{B}$ ions in the BO phase (for $U=3$~eV and hole doping $x=0.2$). Right panel: in plane Ni-O-Ni bond lengths after structural relaxation.}
\label{Fig_1}
\end{figure}

We start by performing structural optimization of the internal atomic positions of $R$NiO$_2$ at different hole dopings using the spin-polarized DFT+U method \cite{Anisimov_1991,Liechtenstein_1995,Dudarev_1998,Giannozzi_2009,Giannozzi_2017}, as implemented in the Quantum ESPRESSO electronic structure package \cite{Giannozzi_2009,Giannozzi_2017}. In order to model a long-range stripe state we adopt the spin and charge stripe patterns as shown in Fig.~\ref{Fig_1}. In these calculations the lattice shape and the lattice parameters $a$ and $c$ were fixed to the experimental values (space group $P4/mmm$, lattice parameters $a=3.91$~\AA\ and $c=3.37$~\AA) \cite{Li_2019} and the computations were performed within the $3\sqrt{2}a\times\sqrt{2}a\times c$ supercell structure (similar to the procedure of Ref.~\onlinecite{Botana_2016}). We use different effective Hubbard $U$ values, starting from the non-interacting DFT case ($U=0$ eV), up to $U_\mathrm{eff}=0$--5~eV, which is typical for the electronic structure studies of nickelates \cite{Werner_2020,Lechermann_2020a,Karp_2020a,Karp_2020b,Lechermann_2020b,
Wang_2020,Nomura_2020,Si_2020,Leonov_2020,Ryee_2020,Lechermann_2021,Wan_2021,
Leonov_2021,Lechermann_2022,Malyi_2022}. Following the literature, to avoid the numerical instabilities arising from the rare-earth $4f$ electrons, we focus on La$^{3+}$ ion as the $R$ ion, exploring the effects of hole doping on the electronic structure of $R$NiO$_2$ within a rigid-band approximation within DFT.

Upon structural optimization of hole-doped $R$NiO$_2$ with the spin and charge stripe pattern depicted in Fig.~\ref{Fig_1} (top) we obtain a remarkable distortion of the Ni-O distances in the Ni-O plane of $R$NiO$_2$, with a significant deviation of the Ni-O bond length from that in the parent undistorted compound (with the Ni-O bond length of $\sim$1.955 \AA). In fact, the difference in the Ni-O bond length of $\sim$0.054-0.075 \AA\ (for different bonds) at $x=0.2$ [see the right panel of Fig.~\ref{Fig_1} (top)] is compatible with the average bond length difference in other charge-disproportionated systems, such as perovskite nickelates \cite{Torrance_1992,Garcia_1992} and iron-based oxides \cite{Wright_2001,Wright_2002,Woodward_2000,Takeda_2000,Leonov_2022,
Greenberg_2018,Ovsyannikov_2016,Layek_2022}. We note that even in the undoped case, $R$NiO$_2$ with $x=0$, a bond length difference is robust, about 0.04 \AA, increasing to 0.07 \AA\ upon hole doping $x=0.4$. Moreover, no sizable buckling of the Ni-O plane is found and the Ni-O plane remains (nearly) flat. 

Our calculations suggest the formations of the charge-disproportionation state with a robust Ni-O bond length difference: ``contracted'' around the nonmagnetic Ni$^{2+}$ ions, Ni$_\mathrm{A}$ in Fig.~\ref{Fig_1} (with a square planar coordination with oxygen ions and Ni-O bond length of $\sim$1.912 \AA) and ``expanded'' around the Ni$^{+}$ $S=1/2$ ions, Ni$_\mathrm{B}$ with $\sim$1.966-1.987 \AA\ (all the numbers are given for $U=3$~eV and hole doping $x=0.2$). The Ni$^+$ ions are seen to be shifted from the center of the planar NiO$_4$ placket to the neighboring Ni$^{+}$ ions row. In addition, we find a remarkable deviation of the Ni-Ni distances (along the Ni-O-Ni path) from that in the undistorted $R$NiO$_2$ ($3.91$ \AA). The Ni-Ni distances are 3.899 \AA\ between the Ni$^{2+}$ and Ni$^{+}$ ions and 3.933 \AA\ between the Ni$^{+}$ ions, resulting in a superstructure modulation with a periodicity of $ 3 \times a$ along the Ni-O bonds (the same behavior is also seen on the $R$-ion sublattice, with the alternating nearest-neighbour $R$-$R$ ion distancies of $\sim$3.917 and 3.895 \AA). This behavior seems agree with the recent Ni $L_3$ RIXS experiments that reveal the formation of a broken translational symmetry state in $R$NiO$_2$ with a wave vector near to $(0.33,0)$ r.l.u., along the Ni-O bond direction \cite{Rossi_2021,Tam_2021,Krieger_2021}. We note that the microscopic origin of this behavior still remains controversial, and one of the possible microscopic explanations is the formation of a charge-density-wave order with a wave vector near to $(0.33,0)$ r.l.u. \cite{Rossi_2021,Tam_2021,Krieger_2021,Chen_2022}.

\begin{figure}[tbp!]
\centerline{\includegraphics[width=0.5\textwidth,clip=true]{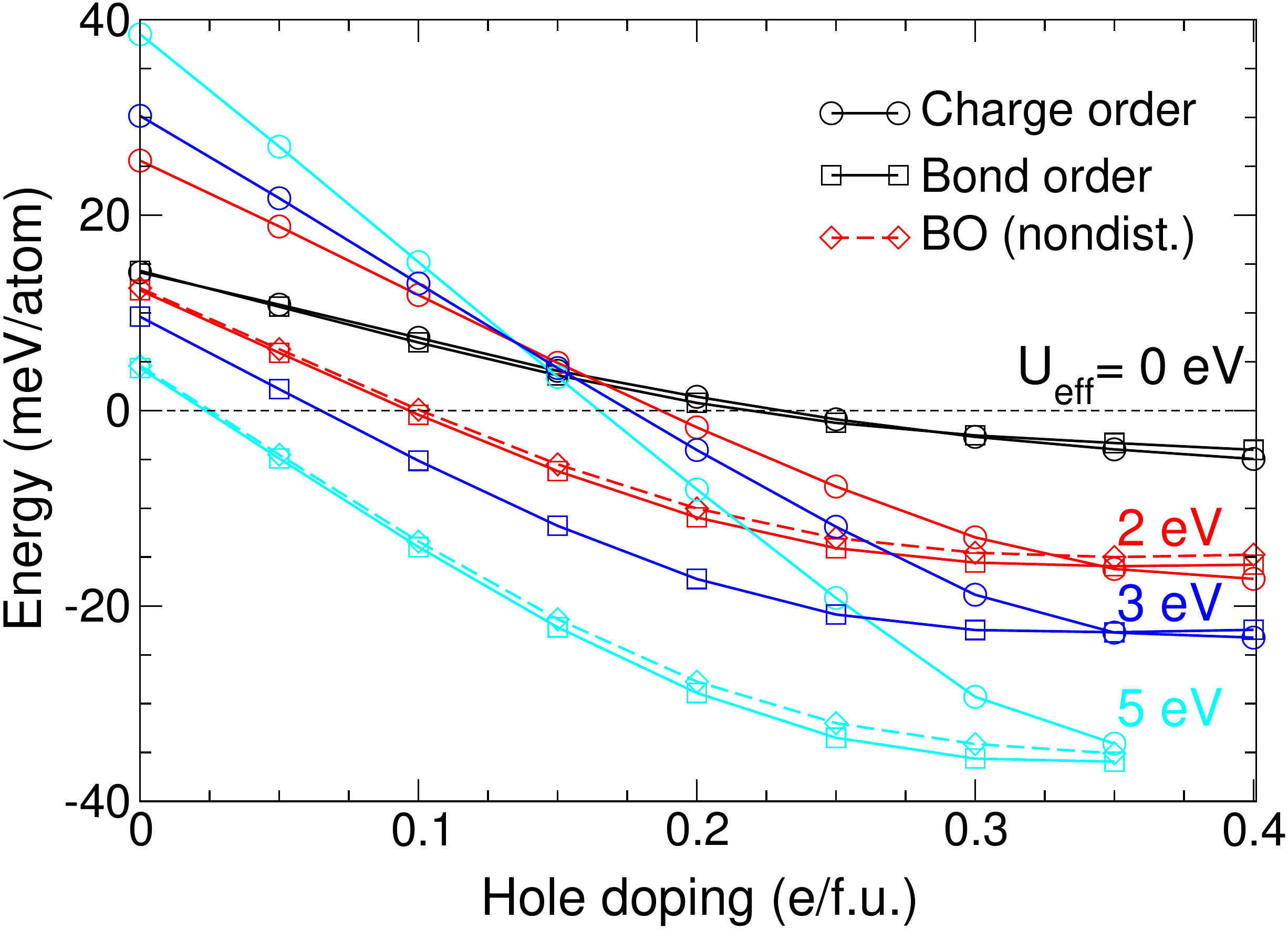}}
\caption{Total energy difference of the charge-ordered and bond-ordered phases of AFM $R$NiO$_2$ evaluated with respect to the $C$-type (110) AFM state (taken as zero energy) using DFT+U as a function of hole doping for different Hubbard interaction values $U_\mathrm{eff}$.}
\label{Fig_2}
\end{figure}

Our calculations (for $U=3$~eV and hole doping $x=0.2$) yield a striped pattern of two Ni$^{+}$ $S=1/2$ (Ni$_\mathrm{B}$ sites with a Ni $3d$ spin moment of 0.68$\mu_\mathrm{B}$) rows followed by one nonmagnetic Ni$^{2+}$ $S=0$ (Ni$_\mathrm{A}$) row, with orientation at $45^\circ$ to the planar Ni-O bonds [see Fig.~\ref{Fig_1} (top)]. While the calculated total Ni $3d$ occupancies at the Ni A and B sites are nearly same, $\sim$9.07, we observe a robust charge-disproportionation characterized by a $\sim$0.13 charge density difference evaluated at the Ni A and B sites (a difference of the site-projected charges). We note that the same stripe order characterized by the formation of AFM with an antiphase domain boundary of hole stripes (centered at the Ni$^{2+}$ $S=0$ ions), with orientation at $45^\circ$ to the Ni-O bond was previously considered as the ground state of the related hole-doped nickelates (La,Sr)$_2$NiO$_4$ (with Sr $x=1/3$, Ni$^{2.33+}$ ions) \cite{Lee_1997,Yoshizawa_2000} and square-planar La$_4$Ni$_3$O$_8$ with Ni$^{1.33+}$ and La$_3$Ni$_2$O$_6$ with Ni$^{1.5+}$ ions \cite{Botana_2016,Zhanga_2016,Bernal_2019,Zhang_2019,Zhang_2020,Hao_2021}. 

\begin{figure}[tbp!]
\centerline{\includegraphics[width=0.5\textwidth,clip=true]{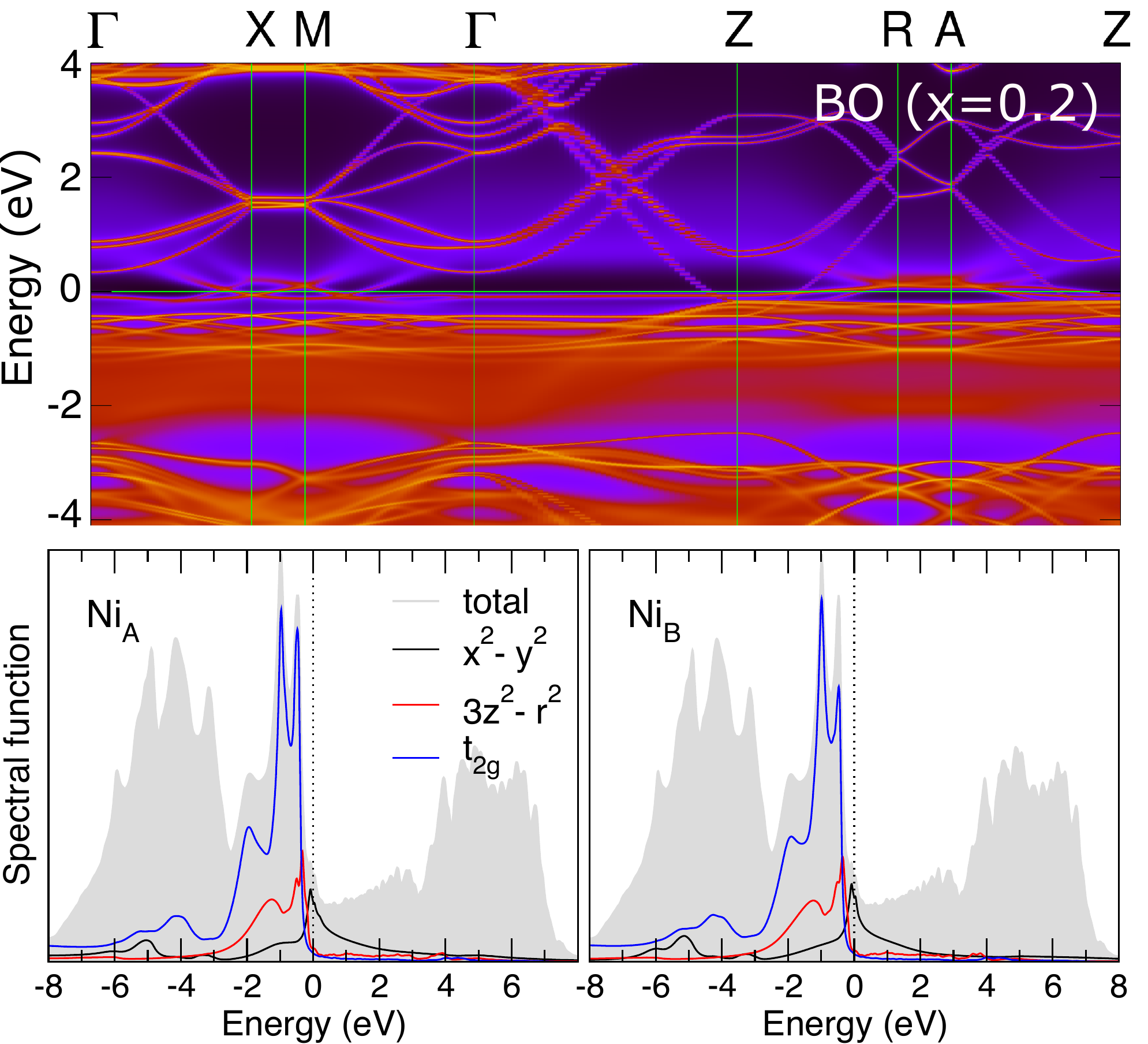}}
\caption{Top: {\bf k}-resolved spectral function of PM $R$NiO$_2$ calculated
by DFT+DMFT for the bond-ordered phase at hole doping $x=0.2$. Bottom: Orbitally resolved spectral functions obtained by DFT+DMFT for the BO PM $R$NiO$_2$ with $x=0.2$.}
\label{Fig_3}
\end{figure}

Next, we perform the DFT+U total energy calculations for the optimized lattice of hole-doped $R$NiO$_2$. Our results obtained for different Hubbard $U$ values are summarized in Fig.~\ref{Fig_2}. While the spin and charge striped phase is found to be thermodynamically unstable at small hole dopings, with the stable $C$-type AFM spin structure, the former becomes stable above a critical doping value of $\sim$0.22 (in the non-interacting DFT case) \cite{Leonov_2020,Leonov_2021}. It is notable that the critical doping depends sensitively on the Hubbard $U$, shifting to about 0.18 for $U=5$ eV. Most interestingly, our calculations suggest the formation of the bond-disproportionated striped phase as depicted in Fig.~\ref{Fig_1} (bottom), which is found to strongly compete with the $C$-type AFM spin state and charge-order stripe phase. This novel striped phase is characterized by a sufficiently smaller Ni-O bond length difference between the Ni$_\mathrm{A}$ (``Ni$^{2+}$'' ions) and Ni$_\mathrm{B}$ (``Ni$^{+}$'' ions) which is of 1.937~\AA\ for Ni$_\mathrm{A}$ (with a regular planar coordination with oxygen ions) and is of 1.958-1.97 \AA\ for Ni$_\mathrm{B}$, respectively (all for $U=3$ eV and hole doping $x=0.2$). As a result, the total Ni $3d$ occupations of the Ni A and B ions are nearly the same, of $\sim$9.05. Moreover, in contrast to the CO stripe state, the Ni A and B site projected charges are the same. Therefore, we term this novel phase as the bond-ordered (BO) stripe phase. In close similarity to the CO phase, the Ni$^+$ ions are seen to be shifted from the center of the planar NiO$_4$ placket to the Ni$^{+}$ ions row, as well as there is a remarkable modulation of the Ni-Ni distances along the Ni-O-Ni path, with a Ni-Ni distance of 3.907 \AA\ between the Ni$_\mathrm{A}$ and Ni$_\mathrm{B}$ ions and 3.916 \AA\ between the Ni$_\mathrm{B}$ ions, which gives a superstructure with the $3 \times a$ modulation of the lattice along the Ni-O bonds. 

In contrast to the charge-ordered phase, for the BO phase our DFT+U calculations give a finite Ni $3d$ spin magnetic moment of 0.56$\mu_\mathrm{B}$ at the Ni$_\mathrm{A}$ sites, the Ni$_\mathrm{B}$ $3d$ spin moment is of 0.66$\mu_\mathrm{B}$, all for $U_\mathrm{eff}=3$ eV and hole doping $x=0.2$. It results in the formation of a concomitant spin-density-wave formed at the Ni$_\mathrm{A}$ sites which holds as an in-phase domain boundary for the AFM state of the Ni$_\mathrm{B}$ S=1/2 ions. As a result, the Ni$_\mathrm{A}$ sites together with the neighboring Ni B $S=1/2$ ions form zigzag ferromagnetic chains in the $ab$ plane, which resembles the unique electronic state of the charge-ordered manganites \cite{Rodriguez_1996,Mori_1998,Mori_1998b,Radaelli_1999,Hemberger_2002,
Loudon_2005}. This picture suggests the possible importance of double exchange mechanism to stabilize the bond-ordered striped phase. In fact, the DFT+U total energy calculations suggest the BO phase to be thermodynamically stable in a broad range of hole dopings, strongly competing with the $C$-type AFM at low and with the charge-ordered striped phase at high doping values. 
Our analysis shows that spin and charge degrees of freedom play a key role in stabilizing the stripe phases, while concomitant lattice displacements according to our calculations give a weak contribution in the total energy (see our results for the BO phase with the undistorted lattice in Fig.~\ref{Fig_2}). This questions strong electron-lattice interactions in hole-doped $R$NiO$_2$.
Moreover, the phase stability of the BO phase is found to depend very sensitively on the choice of the Hubbard $U$ value, being thermodynamically stable at doping level 0.22-0.3 in the non-interacting DFT ($U=0$ eV) and 0.06-0.35 for $U=3$ eV. 

To proceed further we study the electronic structure and quasiparticle band renormalizations of hole-doped $R$NiO$_2$ in the paramagnetic (PM) phase using a fully self-consistent in charge density DFT+DMFT method \cite{Haule_2007,Pourovskii_2007,Leonov_2015,Leonov_2016,Leonov_2020b} implemented with plane-wave pseudopotentials \cite{Leonov_2010,Giannozzi_2009,Giannozzi_2017}. To this end, we adopt the (distorted) crystal structure of the CO and BO phases obtained by performing structural optimization of $R$NiO$_2$ within DFT+U with the Hubbard $U$ value of 5 eV. In particular, we focus on the hole-doped case with $x=0.15$ and 0.2 (near to the optimal doping value) and compute the DFT+DMFT total energies and the electronic structure of all these phases. In the DFT+DMFT calculations we employ the same procedure as it was discussed previously in the context of $R$NiO$_2$ (see Refs.~\onlinecite{Leonov_2020,Leonov_2021}): In DFT+DMFT for the Ni $3d$, La $5d$, and O $2p$ valence states we construct a basis set of atomic-centered Wannier functions within the energy window spanned by these bands \cite{Marzari_2012,Anisimov_2005}. In order to treat the strong on-site Coulomb correlations of the Ni $3d$ electrons within DMFT, we use the average Hubbard parameter $U=6$ eV and Hund's exchange coupling $J=0.95$ eV (i.e., $U_\mathrm{eff} =U-J\sim 5$ eV), with the continuous-time hybridization expansion (segment) quantum Monte Carlo algorithm to solve the realistic many-body problem \cite{Gull_2011}. We use a two-impurity-site DFT+DMFT method in order to treat correlations in the $3d$ bands of the structurally distinct Ni sites in the CO and BO phases.

Our results for the spectral properties (see Fig.~\ref{Fig_3} for the BO $R$NiO$_2$ at $x=0.2$) agree qualitatively with those of the undistorted hole-doped $R$NiO$_2$. We found that the Ni $x^2-y^2$ orbitals are nearly half filled ($\sim$0.55 electrons per spin-orbit for the Ni A and B sites) show a characteristic three-peak structure with a noticeable lower and upper Hubbard subbands and a quasiparticle peak at the Fermi level. The Ni $3z^2-r^2$ orbitals, which are nearly fully occupied (with a spin-orbit occupancy of $\sim$0.839), exhibit a sharp peak in the spectral function at about -0.5 eV below the Fermi level. The latter is associated with the nondispersive electronic states at about -0.5 eV (due to their quasi-2D nature), and is accompanied by a broad subband structure at -1.4 eV. The Ni A and B $3d$ Wannier orbital occupancy difference is small, only of $\sim$0.01, seen as a small occupancy difference of the $x^2 -y^2$ orbitals between the Ni A and B sites. Note that the same value in the CO phase is sufficiently larger, of $\sim$0.024. The instantaneous local moment of Ni ions $\sqrt{\hat{m}^2_z} \sim 1.1 \mu_\mathrm{B}$ is nearly the same for the Ni A and B sites. In agreement with previous DFT+DMFT calculations of infinite-layer $R$NiO$_2$ we found a remarkable orbital-selective renormalization of the partially occupied Ni $x^2 - y^2$ and $3z^2 - r^2$ orbitals \cite{Werner_2020,Lechermann_2020a,Karp_2020a,Karp_2020b,Lechermann_2020b,
Wang_2020,Nomura_2020,Si_2020,Leonov_2020,Ryee_2020,Lechermann_2021,Wan_2021,
Leonov_2021,Lechermann_2022,Malyi_2022}. The Ni $x^2 -y^2$ states show a large quasiparticle mass renormalization of $m^*/m \sim 2.7$ for the Ni A and B sites, while correlation effects in the $3z^2 - r^2$ band are significantly weaker, $\sim$1.4 ($m^*/m$ is derived from the electronic self-energy at the Matsubara frequencies $\omega_n$ as $m^*/m=[1-\partial Im(\Sigma(i\omega_n))/\partial i\omega_n]_{i\omega_n \rightarrow 0}$). Note that at $x=0.15$ a mass renormalization of the Ni $x^2 -y^2$ states in BO $R$NiO$_2$ is somewhat higher, $m^*/m \sim 2.82$, consistent with a previously suggested reduction of the Ni $x^2 -y^2$ band renormalizations upon hole doping \cite{Leonov_2020,Leonov_2021}.

Most interestingly, our DFT+DMFT total energy calculations of PM $R$NiO$_2$ at $T=290$ K and hole doping $x=0.15$ and 0.2 predict a thermodynamic phase stability of the BO phase, with the CO phase being thermodynamically unstable with a total energy difference of $\sim$7 meV/atom. We also note that the BO and the undistorted ($C$-type) phases of PM $R$NiO$_2$ are found to be energetically degenerate within $\sim$1 meV/atom (i.e., within an accuracy of the present calculation). We therefore propose that the BO and the $C$-type phases strongly compete at finite temperatures, while a long-range magnetic order seems to be important for the stabilization of the CO phase. This suggests that dynamical spin fluctuations which are robust at finite temperatures tend to destabilize the CO and BO phases of hole-doped $R$NiO$_2$ at high temperatures.

In fact, our results demonstrate that the $C$-type spin ordered ground state of hole-doped undistorted $R$NiO$_2$ is unstable towards the formation of the spin and charge stripe phases. We observe two striped phases (CO and BO) with different spin-density and charge-/bond-density-wave patterns which are characterized by the emergence of a translational symmetry broken state. The latter is characterized by a sizable variation of the lattice, e.g., of the Ni-Ni distances, with a Ni-O-Ni superstructure with a periodicity of $3 \times a$ along the Ni-O bonds, 
%
%
affecting the electronic structure and exchange interactions in this compound. The latter is seen, e.g., as the formation of the spin-density wave at the Ni$_\mathrm{B}$ sites in the BO phase. While the BO state sets in at low doping value, between 0.06--0.34 (for $U=3$ eV), it is found to be energetically degenerate (or, in other words, strongly competing) with the CO phase at high doping level (e.g., at $x>0.3$), implying possible frustration of the CO and BO stripe states. We therefore speculate that the experimentally detected suppression of the spin and charge stripe state upon doping may stem from a frustration of different stripe states (e.g., CO and BO) at high doping level. Here, we also need to point out that in the present work we have considered only two possible stripe configurations, while other spin and charge stripe arrangements may appear at different or even the same doping. 
In fact, in the present study we do not consider the possible formation of the $(0.33,0)$ charge stripe state in $R$NiO$_2$, which has been addressed in the recent DFT+DMFT study by Chen \emph{et al.} \cite{Chen_2022}.
Our results suggest the emergence of a strong competition between different stripe states on a microscopic level, which affects the electronic structure and superconductivity of this material. Moreover, this raises a question about the possible role of stripe fluctuations to mediate a superconducting state in hole-doped infinite-layer nickelates.


In conclusion, using the DFT+U and DFT+DMFT methods we explore the formation of spin and charge stripes in the Ni-O plane of hole-doped infinite-layer nickelates, $R$NiO$_2$. Our results reveal that the $C$-type spin ordered ground state of hole-doped undistorted $R$NiO$_2$ is unstable towards the formation of the spin and charge stripe phases accompanied by in-plane ``breathing''-type distortions of the crystal structure. We propose two particular candidates, the charge-ordered and bond-disproportionated phases, with a peculiar electronic and spin-state behavior. Our results suggest that the BO phase is thermodynamically stable in a broad range of hole dopings (possibly due to the double-exchange mechanism), strongly competing with the $C$-type AFM at low and with the charge-ordered striped phase at high doping values. This implies a possible frustration of the spin and charge stripe states at high doping value, resulting in the increase of spin and charge stripe fluctuations upon hole doping. 
Our results provide a comprehensive picture of competing charge and spin stripe states, which are key for understanding the anomalous properties of hole-doped layered nickelates. This topic calls for further theoretical and experimental investigations of the intriguing interplay between charge order, AFM, and superconductivity established in the infinite-layer nickelates.

\begin{acknowledgments}
I.V.L. is deeply appreciated to Vasily I. Leonov for his constant help and assistance. The DFT+U and DFT+DMFT electronic structure calculations and structural optimization were supported by the Russian Science Foundation (Project No. 22-22-00926). The theoretical analysis of the electronic structure was supported by the state assignment of Minobrnauki of Russia (theme ``Electron'' No. 122021000039-4).

\end{acknowledgments}

\end{document}